\documentclass[prd,twocolumn,superscriptaddress,altaffilletter,showpacs]{revtex4}
\usepackage{amssymb,amsmath}
\usepackage{comment}
\usepackage{graphicx}
\usepackage{csquotes}
\usepackage{tikz}
\usetikzlibrary{snakes}
\usepackage{lipsum, babel}
\DeclareUnicodeCharacter{2212}{\textendash}

\begin{document}

\title{Schwarzschild black hole and redshift rapidity: \\
A new approach towards measuring cosmic distances}
\author{Mehrab Momennia}
\email{mmomennia@ifuap.buap.mx, momennia1988@gmail.com}
\affiliation{Instituto de F\'{\i}sica, Benem\'erita Universidad Aut\'onoma de Puebla,\\
Apartado Postal J-48, 72570, Puebla, Puebla, Mexico}
\affiliation{Instituto de F\'{\i}sica y Matem\'{a}ticas, Universidad Michoacana de San
Nicol\'as de Hidalgo,\\
Edificio C--3, Ciudad Universitaria, CP 58040, Morelia, Michoac\'{a}n, Mexico}
\author{Pritam Banerjee}
\email{pritam@phy.iitkgp.ac.in, bpritamphy@ gmail.com}
\affiliation{Indian Institute of Technology, Kharagpur, 721302, India}
\author{Alfredo Herrera-Aguilar}
\email{aherrera@ifuap.buap.mx}
\affiliation{Instituto de F\'{\i}sica, Benem\'erita Universidad Aut\'onoma de Puebla,\\
Apartado Postal J-48, 72570, Puebla, Puebla, Mexico}
\author{Ulises Nucamendi}
\email{ulises.nucamendi@umich.mx, unucamendi@gmail.com}
\affiliation{Instituto de F\'{\i}sica y Matem\'{a}ticas, Universidad Michoacana de San
Nicol\'as de Hidalgo,\\
Edificio C--3, Ciudad Universitaria, CP 58040, Morelia, Michoac\'{a}n, Mexico}
\date{\today }

\begin{abstract}
Motivated by recent achievements of a full general relativistic method in
estimating the mass-to-distance ratio of supermassive black holes hosted at
the core of active galactic nuclei, we introduce the new concept
{\it redshift rapidity}
in order to express
the Schwarzschild black hole mass and its distance from the Earth just in
terms of observational quantities. The redshift rapidity is also an
observable relativistic invariant that
represents the evolution of the frequency shift with respect to proper time in the Schwarzschild spacetime. We extract concise and elegant analytic
formulas that allow us to disentangle mass and distance to black holes in the Schwarzschild
background and estimate these parameters separately. This procedure is performed in a completely general relativistic way with the aim of improving
the precision in measuring cosmic distances to astrophysical compact objects.
Our exact formulas are valid on the midline and close to the line of sight, having direct astrophysical applications for megamaser systems, whereas the general relations can be employed in black hole parameter estimation studies. We also computed the frequency shift and the redshift rapidity for emitter eccentric orbits and calculated
their relative error with respect to their numerical exact
value for different eccentricities.

\vskip3mm

\end{abstract}

\maketitle




\section{Introduction}

Black holes first have been found just as mathematical solutions to Einstein
field equations of the general relativity theory by Karl Schwarzschild
whereas their existence in the cosmos has been recently approved through
gravitational wave detections \cite{GW}\ and electromagnetic wave
observations \cite{EHTM87,EHTSgr}. Moreover, investigating the motion of
stars around the center of our galaxy has already provided convincing
evidence for the presence of a supermassive black hole hosted at the core of
the Milky Way galaxy \cite{Ghez,Morris,Eckart,Gillessen}.

Even though the aforementioned methods are fairly accurate in obtaining the black hole parameters using some types of observational data, they are not always applicable to other kinds of existing data from
black holes and black hole environments. Therefore, we have been inventing
and developing an independent general relativistic approach to measure the
black hole and cosmological parameters \cite{PRDhn,PRDbhmn,KdS}. This method
is based on the observational frequency-shifted photons emitted by massive
geodesic particles orbiting central black holes
and has been initially
suggested in \cite{PRDhn}, generalizing previous Keplerian models \cite{Herrnstein2005,Argon2007,Humphreys2013}. Recently, this approach has been developed to
analytically express the mass and spin parameters of the Kerr black hole in
terms of a few directly observational quantities \cite{PRDbhmn}. Besides, it
has been demonstrated that the Kerr black hole in asymptotically de Sitter
background allows us to measure the Hubble constant and black hole
parameters, simultaneously \cite{KdS}.

The initial method of \cite{PRDhn}\ has been employed to obtain the
kinematic redshift and blueshift in terms of black hole parameters for
several black hole solutions, such as Kerr-Newman black holes in de Sitter
spacetime \cite{KNdS}, the Plebanski-Demianski black holes \cite%
{PlebanskiDemianski}, higher dimensional Myers-Perry spacetime \cite%
{MyersPerry}, and regular black holes in nonlinear electrodynamics \cite%
{RegularBH}.\ Furthermore, a similar relation for the kinematic redshift has
been found\ for black holes in modified gravity \cite{MOG}, black holes
coupled to nonlinear electrodynamics \cite{NED}, black holes immersed in a
strong magnetic field \cite{SMF}, and the boson stars \cite{BosonStar}.

These studies have been performed based on the kinematic frequency shift
which is not an observable quantity, unlike the total frequency shift of
photons. Hence, this fact has motivated us to consider the total frequency
shift as a directly observable element and extract concise and elegant
analytic formulas for the mass and spin of the Kerr black hole in terms of
few observable quantities \cite{PRDbhmn}. With the help of this general
relativistic method, the free parameters of polymerized black holes have
been expressed in terms of the total redshift as well \cite{FuZhang}. More
recently, this approach has been extended to general spherically symmetric
spacetimes in order to extract the information of free parameters of black
holes in alternative gravitational theories beyond Einstein gravity \cite%
{Diego}.

From a practical point of view, the mass-to-distance ratio of several
supermassive black holes hosted at the core of the active galactic nuclei
(AGNs) NGC 4258 \cite{ApJL}, TXS-2226-184 \cite{TXS}, and an additional $14$
galaxies \cite{TenAGNs,FiveAGNs} has been estimated by employing this
general relativistic approach. These supermassive black holes enjoy
circularly orbiting water vapor clouds within 
accretion disks that emit redshifted photons
toward a distant observer. This allows us to utilize the general
relativistic method in order to estimate the Mass/Distance ($M/D$) ratio and quantify the
gravitational redshift produced by the spacetime curvature which is a
general relativistic effect.

In the parameter estimation studies of the supermassive black holes
performed in \cite{ApJL,TXS,TenAGNs,FiveAGNs}, just the mass-to-distance ratio $M/D$ 
of these compact objects has been estimated 
with the help of observational redshifted and blueshifted photons
due to the fact that this
ratio is degenerate in this general relativistic formalism. In this paper,
we are going to introduce a new general relativistic invariant observable quantity,
the 
\textit{redshift rapidity}%
, which is the derivative of the redshift with respect to proper time. 
With the aid of the redshift rapidity,
we shall disentangle $M$ and $D$ in the Schwarzschild background to extract
concise and elegant analytic formulas for mass and distance to the black hole just in terms of
observational elements. Thus, these formulas will help us to break the degeneracy
in the $M/D$ ratio of the Schwarzschild black hole and compute the mass and distance to the black hole separately. More recently, the redshift rapidity in the Reissner-Nordstr\"{o}m black hole spacetime has been calculated and employed to extract analytic formulas for the black hole mass and charge as well as its distance from the Earth \cite{PabloRNBH}. We would like to remark that these new formulas enable us to compute the distance to black holes and other astrophysical compact objects in a completely general relativistic framework with the aim of improving the precision in measuring cosmic distances with respect to previous post-Newtonian approaches that compute the angular-diameter distance to several megamaser galaxies like UGC 3789 \cite{MCP2}, NGC 6264 \cite{MCPV}, NGC 5765b \cite{MCPVIII}, NGC 4258 \cite{Humphreys2013,Reid2019}, CGCG 074-064 \cite{MCPXI}, among others. We expect an improvement in precision because general relativity describes the gravitational phenomena in the compact objects' environment in a more complete way than the post-Newtonian approach.

The outline of this paper is as follows. The next section is devoted to a
brief review of our general relativistic formalism in the Schwarzschild
black hole spacetime. Then, we express the mass-to-distance ratio of the
Schwarzschild black hole in terms of observable frequency shifts for three
spacial points of the circular motion of orbiting massive particles, namely, on the
midline and close to the line of sight (LOS). In Sec. III, we define redshift
rapidity as the proper time evolution of the frequency shift in the Schwarzschild
background and employ this observable quantity in order to express the black
hole mass and distance just in terms of observational elements on the
midline and close to the LOS. We further present in Secs. IV and V, respectively, the frequency shift and the redshift rapidity for emitter eccentric orbits and calculate
their relative error with respect to the numerical exact value of these quantities for different eccentricities in Sec. VI. We finish our paper with some
concluding remarks in Sec. VII.

\section{Schwarzschild black hole and the frequency shift}

In order to describe our formalism to measure black hole parameters and cosmic distances, we start
with a brief review on frequency shift formulas of massive probe particles
in the Schwarzschild background based on \cite{PRDbhmn}.\ The Schwarzschild
black hole metric is given by the following line element 
\begin{equation}
ds^{2}=g_{tt}dt^{2}+g_{rr}dr^{2}+g_{\theta \theta }d\theta ^{2}+g_{\varphi
\varphi }d\varphi ^{2},  \label{metric}
\end{equation}%
with the metric components 
\begin{equation}
g_{\mu \nu }=diag\left[ -\left( 1-\frac{2M}{r}\right) ,\left( 1-\frac{2M}{r}%
\right) ^{-1},r^{2},r^{2}\sin ^{2}\theta \right] ,
\end{equation}%
where $M$ is the total mass of the black hole and the event horizon is
located at the Schwarzschild\ radius $r_{+}=2M$.

The massive geodesic particles revolving the Schwarzschild black hole feel
the curvature of spacetime produced by the black hole mass and keep memory
of it. Besides, the observers located on these particles can exchange
redshifted photons that have information about the Schwarzschild black hole
mass in their frequency shift from emission till detection. Thus, the shifts in
the frequency of photons along with the orbital parameters of the emitter
can be used to determine the black hole parameters \cite{ApJL,TXS,TenAGNs,FiveAGNs}.
Therefore, this formalism allows one to compute the black hole parameters in
terms of directly measured observational quantities and the orbital parameters of the emitter \cite{PRDbhmn,FuZhang,Diego}.

Within General Relativity, the frequency of a photon that is
emitted/detected by an emitter/observer with the proper $4$-velocity $%
U_{p}^{\mu }=(U^{t},U^{r},U^{\theta },U^{\varphi })\mid _{p}$ at some
position $x_{p}^{\mu }=(x^{t},x^{r},x^{\theta },x^{\varphi })\mid _{p}$
reads 
\begin{equation}
\omega _{p}=-\left( k_{\mu }U^{\mu }\right) \mid _{p}\,,  \label{freq}
\end{equation}%
where $k_{p}^{\mu }=\left( k^{t},k^{r},k^{\theta },k^{\varphi }\right) \mid
_{p}$ is the $4$-momentum of the photons with the null condition $k_{\mu
}k^{\mu }=0$ and the index ${p}$ refers to either the emission point $%
x_{e}^{\mu }$ or detection point $x_{d}^{\mu }$ of the photons.
Additionally, the $4$-velocity of particles is normalized to unity $U_{\mu
}U^{\mu }=-1$, and when the detector's orbit is located far away from the
emitter-black hole system, $U_{d}^{\mu }$ reduces to 
\begin{equation}
U_{d}^{\mu }=(1,0,0,0)\,.  \label{DisObs}
\end{equation}

The frequency shift of light signals emitted by massive geodesic particles
in equatorial circular orbits revolving the spherically symmetric background
(\ref{metric}) is given by \cite{PRDhn,PRDbhmn} 
\begin{equation}
1+z_{_{Schw}}\!=\frac{\omega _{e}}{\omega _{d}}=\frac{(E_{\gamma
}U^{t}-L_{\gamma }U^{\varphi })\mid _{e}}{(E_{\gamma }U^{t}-L_{\gamma
}U^{\varphi })\mid _{d}},  \label{freq_shift_circular}
\end{equation}%
where the conserved quantities $E_{\gamma }$ and $L_{\gamma }$ stand for the
total energy and axial angular momentum of the photons. By considering Eq. (%
\ref{DisObs}), this relation reduces to%
\begin{equation}
1+z_{_{Schw}}\!=(U^{t}-b_{\gamma }U^{\varphi })\mid _{e},  \label{z}
\end{equation}%
for distant observers and $b_{\gamma }=L_{\gamma }/E_{\gamma }$ is the
deflection of light parameter that takes into account the light bending
generated by the gravitational field in the vicinity of the Schwarzschild
black hole. The non-vanishing components of the $4$-velocity of particles in the equatorial circular motion read \cite{Diego} 
\begin{equation}
U_{e}^{t}=\left. \sqrt{\frac{2}{2g_{tt}-rg_{tt}^{\prime }}}\right\vert
_{r=r_{e}}=\sqrt{\frac{r_{e}}{r_{e}-3M}},  \label{tCompVelo}
\end{equation}%
\begin{equation}
U_{e}^{\varphi }=\pm\left. \frac{1}{r}\sqrt{\frac{rg_{tt}^{\prime }}{%
2g_{tt}-rg_{tt}^{\prime }}}\right\vert _{r=r_{e}}=\pm\frac{1}{r_{e}}\sqrt{\frac{%
M}{r_{e}-3M}},  \label{phiCompVelo}
\end{equation}%
where a prime denotes $\partial _{r}$, $r_{e}$ is the radius of the emitter, and $+/-$ sign refers to counterclockwise/clockwise rotation of the emitter as seen by the observer.

In addition, the $\left(\varphi+\delta \right)$-dependent light bending parameter is
given by [see Eq. (\ref{bGamma}) of the Appendix] 
\begin{eqnarray}
b_{\gamma } &=&\frac{-g_{\varphi \varphi }\sin \left( \varphi+\delta \right) 
}{\sqrt{-g_{tt}}\sqrt{g_{\varphi \varphi }\sin ^{2}\left( \varphi+\delta \right) +r_{e}^{2}g_{rr}\cos ^{2}\left( \varphi+\delta \right) }}  \notag \\
&=&-\frac{r_{e}^{3/2}\sin \left( \varphi+\delta \right) }{\sqrt{r_{e}-2M\sin
^{2}\left( \varphi+\delta \right) }},  \label{lbp}
\end{eqnarray}%
where $\varphi$ is the unobservable azimuthal angle ranging $0\leq \varphi \leq 2\pi $ and $\delta $\ is the aperture angle of the telescope (angular distance) that is a measurable quantity  (see Fig. \ref{PhiDeltaFig} of the Appendix). Both $\varphi$ and $\delta$ are positive (negative) when measured counterclockwise (clockwise) with respect to the LOS.

Now, by substituting Eqs. (\ref{tCompVelo})-(%
\ref{lbp})\ into the redshift formula (\ref{z}), we can find the frequency
shift of photons emitted by massive geodesic particles on an arbitrary point
of equatorial circular orbits in the Schwarzschild background as follows%
\begin{eqnarray}
&&1+z_{Schw}\!\!=\sqrt{\frac{r_{e}}{r_{e}-3M}}\times  \notag \\
&&\left( 1\pm\sqrt{\frac{M}{r_{e}-2M\sin ^{2}\left( \varphi+\delta \right) }}\sin \left( \varphi+\delta \right) \right) ,  \label{zSchw}
\end{eqnarray}%
with the contribution of the gravitational redshift $z_{g}$\ and the
kinematic redshift $z_{kin_\pm}$ as%
\begin{equation}
z_{g}=-1+\sqrt{\frac{r_{e}}{r_{e}-3M}},
\end{equation}%
\begin{equation}
z_{kin_\pm}=\pm\sqrt{\frac{r_{e}}{r_{e}-3M}}\sqrt{\frac{M}{r_{e}-2M\sin
^{2}\left( \varphi+\delta \right) }}\sin \left( \varphi+\delta \right) ,
\label{zkinpm}
\end{equation}%
satisfying $z_{_{Schw}}=z_{_{g}}+z_{kin_\pm}$. In what follows we shall restrict ourselves to the counterclockwise motion of the emitter as seen by a distant observer, using the $+$ sign in Eqs. (\ref{phiCompVelo}), (\ref{zSchw}), and (\ref{zkinpm}) in order to avoid any ambiguity.

In the Newtonian limit $M/r_{e} \rightarrow 0$,  
the frequency shift in the Schwarzschild spacetime (\ref{zSchw}) reduces to the projection of the Keplerian velocity of a particle in circular motion on the LOS in the following way

\begin{equation}
z_{N}=\sqrt{\frac{M}{r_{e}}}\sin (\varphi +\delta )+\mathcal{O}\left( \frac{M%
}{r_{e}}\right) ,
\end{equation}
as we expected. Besides, in this case the azimuthal angle $%
\varphi $ is a function of the
observable $\delta $\ for an arbitrary point on
the circular orbit as follows [see Eq. (\ref{deltaVSphi}) of the Appendix]%
%
\begin{equation}
\delta =\arccos \left( \frac{D-r_{e}\cos \varphi }{\sqrt{%
D^{2}+r_{e}^{2}-2Dr_{e}\cos \varphi }}\right) ,  \label{deltaPhiRel}
\end{equation}%
where $D$\ is the radial distance between the black hole center and the observer.

\begin{figure*}[t]
\centering
\includegraphics[scale=.95]{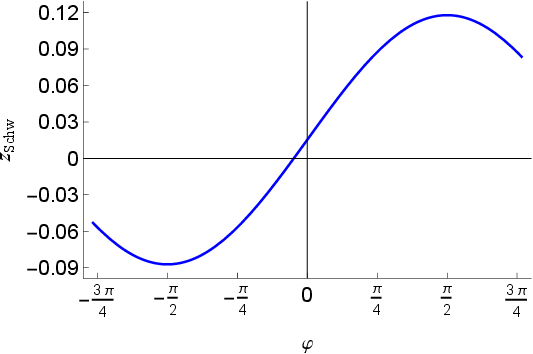}
\includegraphics[scale=.95]{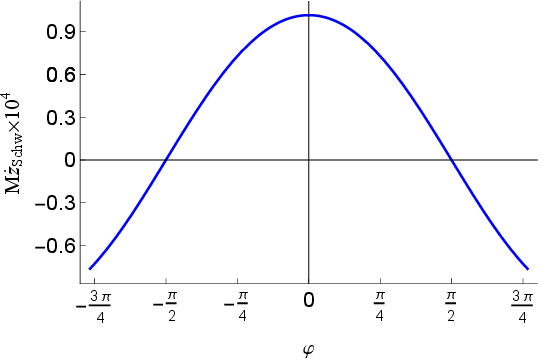}
\caption{The frequency shift $z_{Schw}$ (left panel) and the redshift rapidity $\dot{z}_{Schw}$ (right panel) versus the azimuthal angle in the Schwarzschild background for $r_{e}=10^2M$ and $D=10^5M$. The shift in the frequency is maximal close to the midline where $\varphi\approx\pm \pi /2$, whereas the maximum redshift rapidity is at the LOS $\varphi=0$. In Eqs. (\ref{zSchw}) and (\ref{RedRap}), we replaced $\delta$ with Eq. (\ref{deltaPhiRel}) to plot these curves.}
\label{zDotFig}
\end{figure*}

Fig. \ref{zDotFig} illustrates the redshift formula (\ref{zSchw}) versus the azimuthal angle $\varphi$. This figure shows how the frequency shift in the Schwarzschild spacetime changes with the motion of the massive test particle and the non-vanishing value of the frequency shift at the LOS indicates the gravitational redshift. 
Besides, on the midline where $\varphi=\pm \pi /2$, the total frequency shift is maximal, hence easier to be identified observationally. The range of azimuthal angle used to plot Fig. \ref{zDotFig} is determined to respect the assumption underlying Eq. (\ref{deltaPhiRel}), which considers that light bending is minimal. 

In order to express the mass-to-distance ratio of the Schwarzschild black
hole in terms of observational quantities, it is convenient to consider two
special important cases useful for describing the frequency shift of photon sources within accretion disks revolving supermassive black holes hosted at the core of AGNs.

\subsection{Midline case}

The first case is related to highly shifted frequency photons emitted from
particles located at the midline where $\varphi =\pm \pi /2$. The position
vector of these particles with respect to the black hole center is
orthogonal to the observer's LOS. Hence, by substituting $\varphi
=\pm \pi /2$\ in the frequency shift formula (\ref{zSchw}), one can find the
highly redshifted ($\varphi =+\pi /2$) and blueshifted ($\varphi =-\pi /2$)
photons as below%
\begin{equation}
1+z_{_{Schw_{1,2}}}^{m}\!\!\!=\sqrt{\frac{r_{e}}{r_{e}-3M}}\left( 1\pm \sqrt{%
\frac{M}{r_{e}-2M\cos ^{2}\delta _{m}}}\cos \delta _{m}\right)
\end{equation}%
where the index \textquotedblleft $m$"\ means the aperture angle $\delta $\
should be measured on the midline and now the plus (minus) sign refers to the
redshifted (blueshifted) photons denoted by $z_{_{Schw_{1}}}^{m}$ ($%
z_{_{Schw_{2}}}^{m}$) on the midline. By multiplying $%
R_{m}:=1+z_{_{Schw_{1}}}^{m}$\ and $B_{m}:=1+z_{_{Schw_{2}}}^{m}$, it is
straightforward to obtain the mass-to-distance ratio of the Schwarzschild
black hole in terms of observational quantities on the midline as follows%
\begin{equation}
\frac{M}{D\delta _{m}}=\frac{R_{m}B_{m}-1}{2R_{m}B_{m}}-\frac{\left(
1-R_{m}B_{m}\right) \left( 3-4R_{m}B_{m}\right) }{6R_{m}B_{m}(3-R_{m}B_{m})}%
\delta _{m}^{2},  \label{MDratio}
\end{equation}%
where we used the approximation (\ref{AppMid}) where $D\gg r_{e}$\ is the
radial distance between the black hole and the detector. Additionally, we
considered the limit $\delta _{m}\rightarrow 0$\ in consistency with the
condition (\ref{DisObs}) for distant observers\ as well. As we can see from
Eq. (\ref{MDratio}), the mass-to-distance ratio of the Schwarzschild black
hole is expressed in terms of directly observational elements $\left\{
R_{m},B_{m},\delta _{m}\right\} $.

\subsection{Line of sight case}

The second case is describing the frequency shift of photons emitted close to the LOS where $\varphi \rightarrow 0$. Hence, by substituting $\varphi =\varphi
_{s}\sim 0$\ in the frequency shift formula (\ref{zSchw}), we find the expressions for slightly redshifted and slightly blueshifted photons as below%
\begin{equation}
1+z_{_{Schw_{1,2}}}^{s}\!=\sqrt{\frac{r_{e}}{r_{e}-3M}}\pm \sqrt{\frac{M}{%
r_{e}-3M}}\left( \varphi _{s}+\delta _{s}\right) ,
\end{equation}%
where the index \textquotedblleft $s$"\ means the measurement\ should be
performed for systemic particles (particles close to the LOS) and
we applied the limit $\delta _{s}\rightarrow 0$\ simultaneously. Additionally, we
have just the gravitational redshift $z_{g}$ exactly at the LOS
with $\varphi _{s}=0=\delta _{s}$. In this relation, the angles $\delta _{s}$%
\ and $\varphi _{s}$\ should be measured close to the LOS, and the
plus (minus) sign refers to the redshifted (blueshifted) photons close to the LOS denoted by $z_{_{Schw_{1}}}^{s}$ ($z_{_{Schw_{2}}}^{s}$), respectively.
Now, by multiplying $R_{s}:=1+z_{_{Schw_{1}}}^{s}$\ and $%
B_{s}:=1+z_{_{Schw_{2}}}^{s}$, we can obtain the mass-to-distance ratio of
the Schwarzschild black hole close to the LOS as follows%
\begin{equation}
\frac{M\varphi _{s}}{D\delta _{s}}=\frac{R_{s}B_{s}-1}{3R_{s}B_{s}}+\mathcal{%
O}\left( \varphi _{s}^{2}\right) ,  \label{MDratioLOS}
\end{equation}%
in which we used the approximation $r_{e}\sim D\delta _{s}/\varphi _{s}$\
from Eq. (\ref{AppLOS}) valid close to the LOS and considered the
limits $\left\{ \delta _{s}\rightarrow 0,\varphi _{s}\rightarrow 0\right\} $
simultaneously to derive this equation. Furthermore, this relation is a
function of a set of purely observational quantities $\left\{ R_{s},B_{s}\right\} $.

If we combine the result (\ref{MDratioLOS}), valid for close to the LOS azimuthal angles, with the midline relation (\ref{MDratio}), we can express $\varphi_s$ in terms of purely observational quantities  $\left\{
R_{m},B_{m},\delta _{m};R_{s},B_{s},\delta _{s}\right\} $.

\section{Disentangling mass and distance: The role of redshift rapidity}

The mass-to-distance ratio formulas (\ref{MDratio}) and (\ref{MDratioLOS})
of the Schwarzschild black hole can be employed to estimate the $M/D$\ ratio
of real astrophysical systems as it was accomplished for the central
supermassive black holes of $17$ galaxies in \cite{ApJL,TXS,TenAGNs,FiveAGNs}%
. As the next step in this direction, we are going to disentangle the mass $%
M $\ and the distance to the black hole $D$ to obtain closed formulas for
each parameter in terms of the set of observational elements $\left\{
R_{m},B_{m},\delta _{m};R_{s},B_{s},\delta _{s}\right\} $. 
In order to achieve this aim, we define the frequency shift rapidity, henceforth called {\it
redshift rapidity}, as the proper time evolution of the frequency shift $%
z_{Schw}$ (\ref{z}) in the Schwarzschild background as follows 
\begin{equation}
\dot{z}_{Schw,e}=\frac{dz_{Schw}}{d\tau }=\frac{d}{d\tau }(U_{e}^{t}-b_{\gamma
}U_{e}^{\varphi }),  \label{RR0}
\end{equation}%
which is the redshift rapidity at the emission point and constitutes a general relativistic invariant quantity since the proper time  also shares this property. Notwithstanding, we should note that the redshift rapidity needs to be measured from the Earth. Therefore, by making use of the chain rule, we rewrite Eq. (\ref{RR0}) at the observer position as
\begin{equation}
\dot{z}_{Schw}=\frac{dz_{Schw}}{dt }=\frac{d\tau}{dt }\frac{dz_{Schw}}{d\tau }=\frac{1}{U_{e}^{t}}\frac{d}{d\tau }(U_{e}^{t}-b_{\gamma
}U_{e}^{\varphi }),  \label{RR}
\end{equation}
which is an observable quantity that we measure here on the Earth. For massive geodesic particles
circularly orbiting the black hole in the equatorial plane, the redshift
rapidity (\ref{RR}) reduces to 
\begin{equation}
\dot{z}_{Schw}=-\frac{db_{\gamma }}{d\tau }\ \frac{U_{e}^{\varphi }}{%
U_{e}^{t}},
\end{equation}%
since the $t$-component $U_{e}^{t}$\ (\ref{tCompVelo}) and $\varphi $%
-component $U_{e}^{\varphi }$\ (\ref{phiCompVelo}) of the $4$-velocity are
constant quantities for circular orbits whereas the light bending parameter (%
\ref{lbp}) is time dependent through $\delta $\ and $\varphi $. By employing
the chain rule, we have%
\begin{equation}
\dot{z}_{Schw}=-\left( \frac{\partial b_{\gamma }}{\partial \varphi }+\frac{%
\partial b_{\gamma }}{\partial \delta }\frac{\partial \delta }{\partial
\varphi }\right) \frac{\left( U_{e}^{\varphi }\right) ^{2}}{U_{e}^{t}},
\end{equation}%
where we used $U_{e}^{\varphi }=\left. \frac{d\varphi }{d\tau }\right\vert
_{r=r_{e}}$. Now, we perform the aforementioned derivatives\ by taking into
account the light bending parameter (\ref{lbp}) and the $\delta \left( \varphi
\right) $-relation (\ref{deltaPhiRel})\ to obtain the redshift rapidity for
an arbitrary point on the circular orbit as below%
\begin{eqnarray}
&&\dot{z}_{Schw}\!\!=\frac{1}{\left[ r_{e}-2M\sin ^{2}(\varphi +\text{$%
\delta $})\right] ^{\frac{3}{2}}}\times   \notag \\
&&\frac{DM\left( D-r_{e}\cos \varphi \right) \cos (\varphi +\text{$\delta $})%
}{\sqrt{r_{e}-3M}\left( D^{2}+r_{e}^{2}-2Dr_{e}\cos \varphi \right) }\!.
\label{RedRap}
\end{eqnarray}

Fig. \ref{zDotFig} shows the behavior of the redshift rapidity versus the azimuthal angle $\varphi$. As one can see from this figure, the redshift rapidity at the LOS is maximal, hence it is
easier to measure this quantity at $\varphi\approx 0$.

It is worth noting that the redshift rapidity (\ref{RedRap}) reduces to the projection of the Keplerian acceleration of a particle in circular motion on the LOS in the Newtonian limit $M/r_{e} \rightarrow 0$ and large distances $r_{e}/D \rightarrow 0$
\begin{equation}
\dot{z}_{N}=\frac{M}{r_{e}^{2}}\cos (\varphi +\delta )+\mathcal{O}\left( 
\frac{M^{2}}{r_{e}^{3}}\right) ,
\end{equation}
as it should be.

As the final step, based on the emitter-black hole system configuration and
observational data availability, we can employ the redshift rapidity (\ref%
{RedRap}) as well as either Eq. (\ref{MDratio}) or Eq. (\ref{MDratioLOS}) to
disentangle the Schwarzschild black hole mass $M$\ and the distance to the
black hole $D$. Here, we are going to disentangle $M$\ and $D$\ for two
important cases describing astrophysical photon sources within 
accretion disks circularly orbiting central
supermassive black holes at the core of AGNs.

\subsection{Midline case}

The first important case is related to the emitters on the midline where
their emitted photons are highly redshifted/blueshifted. The
mass-to-distance ratio of the central black hole in terms of these frequency
shifted photons is given in Eq. (\ref{MDratio}). In order to find the
redshift rapidity of the same redshifted particles, we substitute the
approximation (\ref{AppMid}) for $r_{e}$ in the redshift rapidity relation (%
\ref{RedRap}) and consider the limit $\delta _{m}\rightarrow 0$\ to get%
\begin{eqnarray}
\dot{z}_{Schw}^{m} &=&\frac{\frac{M}{D\delta _{m}}}{D\sqrt{1-3\frac{M}{%
D\delta _{m}}}\left( 1-2\frac{M}{D\delta _{m}}\right) ^{\frac{3}{2}}}\times 
\notag \\
&&\left( 1+\frac{11-28\frac{M}{D\delta _{m}}-12\left( \frac{M}{D\delta _{m}}%
\right) ^{2}}{6\left( 1-3\frac{M}{D\delta _{m}}\right) \left( 1-2\frac{M}{%
D\delta _{m}}\right) }\delta _{m}^{2}\right),  \label{RedRapMid}
\end{eqnarray}%
up to second order in $\delta _{m}$. Finally, one can find the explicit form
of the Schwarzschild black hole mass\ and distance to the black hole by
solving Eqs. (\ref{MDratio}) and (\ref{RedRapMid}) as below%
\begin{equation}
M=\frac{\left( R_{m}B_{m}-1\right) ^{2}}{2\dot{z}_{Schw}^{m}\sqrt{%
6-2R_{m}B_{m}}}\delta _{m}+\mathcal{O}\left( \delta _{m}^{3}\right) ,
\label{Mass}
\end{equation}%
\begin{eqnarray}
&&D\!\!=\frac{R_{m}B_{m}\left( R_{m}B_{m}-1\right) }{\dot{z}_{Schw}^{m}\sqrt{%
6-2R_{m}B_{m}}}\times   \label{Distance} \\
&&\left( 1+\frac{R_{m}B_{m}(3R_{m}B_{m}-7)(8R_{m}B_{m}-27)-36}{%
6(R_{m}B_{m}-3)^{2}}\delta _{m}^{2}\right) \!,  \notag
\end{eqnarray}%
fully expressed in terms of observational quantities $\left\{ R_{m},B_{m},%
\dot{z}_{Schw}^{m},\delta _{m}\right\} $\ that should be measured on the
midline.

\subsection{Line of sight case}

The second important case belongs to the photon sources that lie close to the LOS where their emitted photons are slightly
redshifted/blueshifted but have maximum redshift rapidity. The
mass-to-distance ratio of the central black hole in terms of these frequency
shifted photons is given by Eq. (\ref{MDratioLOS}). Thus, we substitute $%
r_{e}=D\delta _{s}/\varphi _{s}$ from the approximation (\ref{AppLOS}) into
the redshift rapidity formula (\ref{RedRap}) and apply the limits $\left\{ \delta
_{s}\rightarrow 0,\varphi _{s}\rightarrow 0\right\} $ to get%
\begin{equation}
\dot{z}_{Schw}^{s}=\frac{1}{M}\left( 1-3\frac{M\varphi _{s}}{D\delta _{s}}%
\right) ^{-\frac{1}{2}}\left( \frac{M\varphi _{s}}{D\delta _{s}}\right) ^{2},
\end{equation}%
for the photon sources close to the LOS. Now, by substituting Eq.
(\ref{MDratioLOS}) in this relation, we can find the Schwarzschild black
hole mass as follows%
\begin{equation}
M=\frac{\left( R_{s}B_{s}-1\right) ^{2}}{9\dot{z}_{Schw}^{s}\left(
R_{s}B_{s}\right) ^{\frac{3}{2}}},
\end{equation}%
fully in terms of observational quantities $\left\{ R_{s},B_{s},\dot{z}%
_{Schw}^{s}\right\} $\ that should be measured close to the LOS.
As the next step, we take advantage of this relation and Eq. (\ref{MDratio})
in order to combine observations from both midline and LOS to get
the distance to the black hole as 
\begin{equation}
D=\frac{\left( R_{s}B_{s}-1\right) ^{2}}{9\dot{z}_{Schw}^{s}\left(
R_{s}B_{s}\right) ^{\frac{3}{2}}}\frac{2R_{m}B_{m}}{\delta _{m}\left(
R_{m}B_{m}-1\right) },
\end{equation}%
which is a significant relation because it contains information on both
high-frequency shifted\ particles on the midline $\left\{
R_{m},B_{m}\right\} $ and maximum redshift rapidity close to the LOS\ $\dot{z}_{Schw}^{s}$ (see Fig. \ref{zDotFig} and related discussion). From an observational point of view, this formula is important due to the fact that it is easier to identify frequency shifts on the midline and the redshift rapidity close to the LOS.

\section{Frequency shift in case of elliptical orbit of the emitter}

So far, we have discussed the redshift and redshift rapidity, considering that the emitter is in a circular orbit. While circularity is a very good assumption, the emitter orbit may differ from the circularity and acquire a small eccentricity in real astrophysical situations. Keeping this in mind, we, in this section, undertake the above analysis using an elliptical emitter orbit. Next, we set the eccentricity to be small and express the relations up to the first order in eccentricity.

In the Schwarzschild metric, we can consider the emitter to be orbiting in the equatorial plane with 4-velocity $U_e^\mu = \{U_e^t, U_e^r, 0, U_e^\varphi \}$ without loss of generality. From the 4-velocity normalization of timelike particles in the Schwarzschild background, we can write
\begin{equation}
(U_e^r)^2 = E^2 - (r_e-2 M) \left(\frac{1}{r_e} + \frac{L_z^2}{r_e^3}\right),\label{4vel_norm1}
\end{equation}
where $E$ and $L_z$ are constants of motion, namely, the orbital energy and $z$-component of the orbital angular momentum of the emitter particle, respectively, defined as,
\begin{equation}
E = \left(1-\frac{2M}{r_e}\right) U^t,~~~ L_z = r_e^2 U^\varphi.
\label{E_Lz}
\end{equation}
We can rewrite Eq. (\ref{4vel_norm1}) as
\begin{align}
(U_e^r)^2 &= 2 \epsilon - 2\nu(r_e),
\label{4vel_norm2}
\end{align}
where
\begin{equation}
2\epsilon = E^2 -1, ~~~ 2\nu(r_e) = \frac{L_z^2}{r_e^2}\left(1- \frac{2M}{r_e}\right) - \frac{M}{r_e}.
\label{epsilon_nu}
\end{equation}

Throughout the elliptical orbit, $E$ and $L_z$ remain fixed. We express the orbital radius of the emitter as 
\begin{equation}
r_e = \frac{p}{1+e \cos\chi}, \label{eqn:conicsection}
\end{equation}
where $p$ is the semi-latus rectum, $e$ is the eccentricity of the elliptical orbit, and $\chi $ is the angular position of the emitter measured from the semi-major axis such that $\chi = 0$ stands for the pericenter position. Using the fact that at the pericenter $r_e=r_p$ and apocenter $r_e=r_a$ of the elliptical orbit, $U_e^r(r_p) = 0 = U_e^r(r_a)$, we can use Eq. (\ref{4vel_norm2}) to find $E$ and $L_z$ as
\begin{align}
    E &= \sqrt{\frac{p^2 - 4 p M + 4 M^2 (1 - e^2)}{p\left(p-M(e^2 + 3)\right)}}, \label{eqn:E}\\ 
    L_z  &= \pm \sqrt{\frac{M p^2}{p - M (e^2 + 3)}}. \label{eqn:Lz}
\end{align}

Here, according to our convention, the plus (minus) sign in Eq. (\ref{eqn:Lz}) stands for the counterclockwise (clockwise) rotation of the emitter as seen by the observer. Now, putting Eqs. (\ref{eqn:E}) and (\ref{eqn:Lz}) in (\ref{4vel_norm2})-(\ref{epsilon_nu}) we get
\begin{equation}
    U_e^r = \pm ~ e \sin\chi \left[ \frac{p M-2 M^2 \left(3 + e \cos\chi\right)}{p^2 - p M (e^2 + 3)} \right]^{1/2}. \label{eqn:ur}
\end{equation}

Note that $U_e^r$ can take the plus (minus) sign for the emitter's counterclockwise (clockwise) motion. From Eqs. (\ref{E_Lz}), (\ref{eqn:E}), and (\ref{eqn:Lz}), we can further obtain the other two components of the 4-velocity. The sign of $U_e^\varphi$ is the same as that of $L_z$. It follows our convention that $\varphi$ and $\delta$ are again positive (negative) when measured counterclockwise (clockwise) from the LOS.

\begin{figure*}[ht]
\centering
\includegraphics[scale=.65]{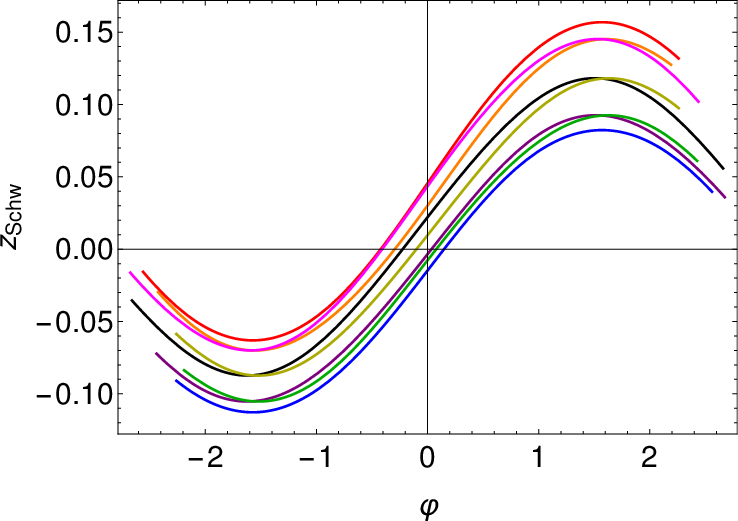}
\hspace{0.5cm}
\includegraphics[scale=.65]{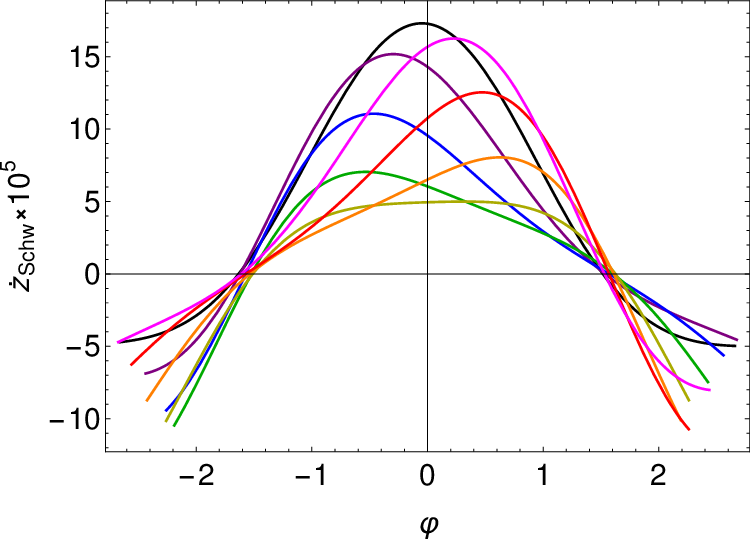}

\includegraphics[scale=.65]{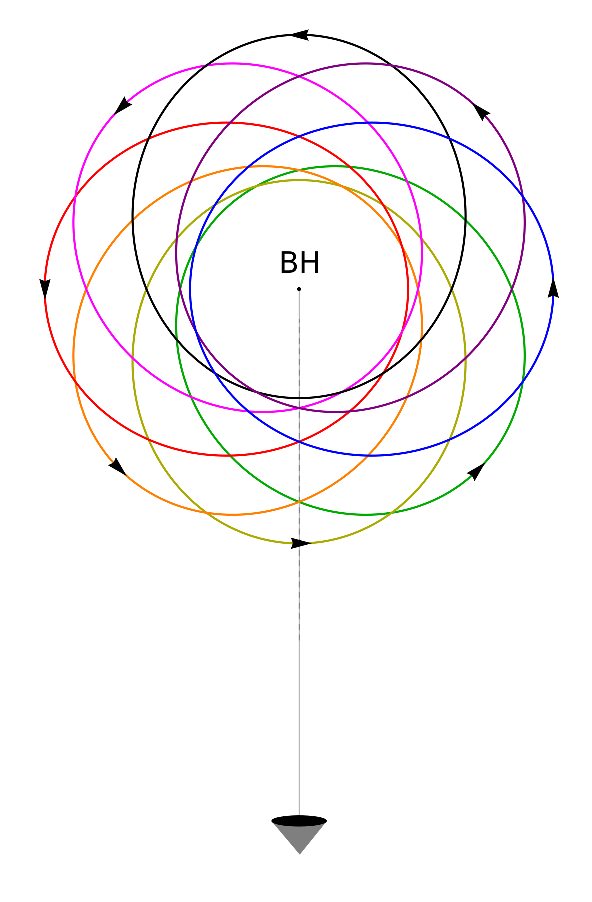}
\hspace{1.5cm}
\includegraphics[scale=.35]{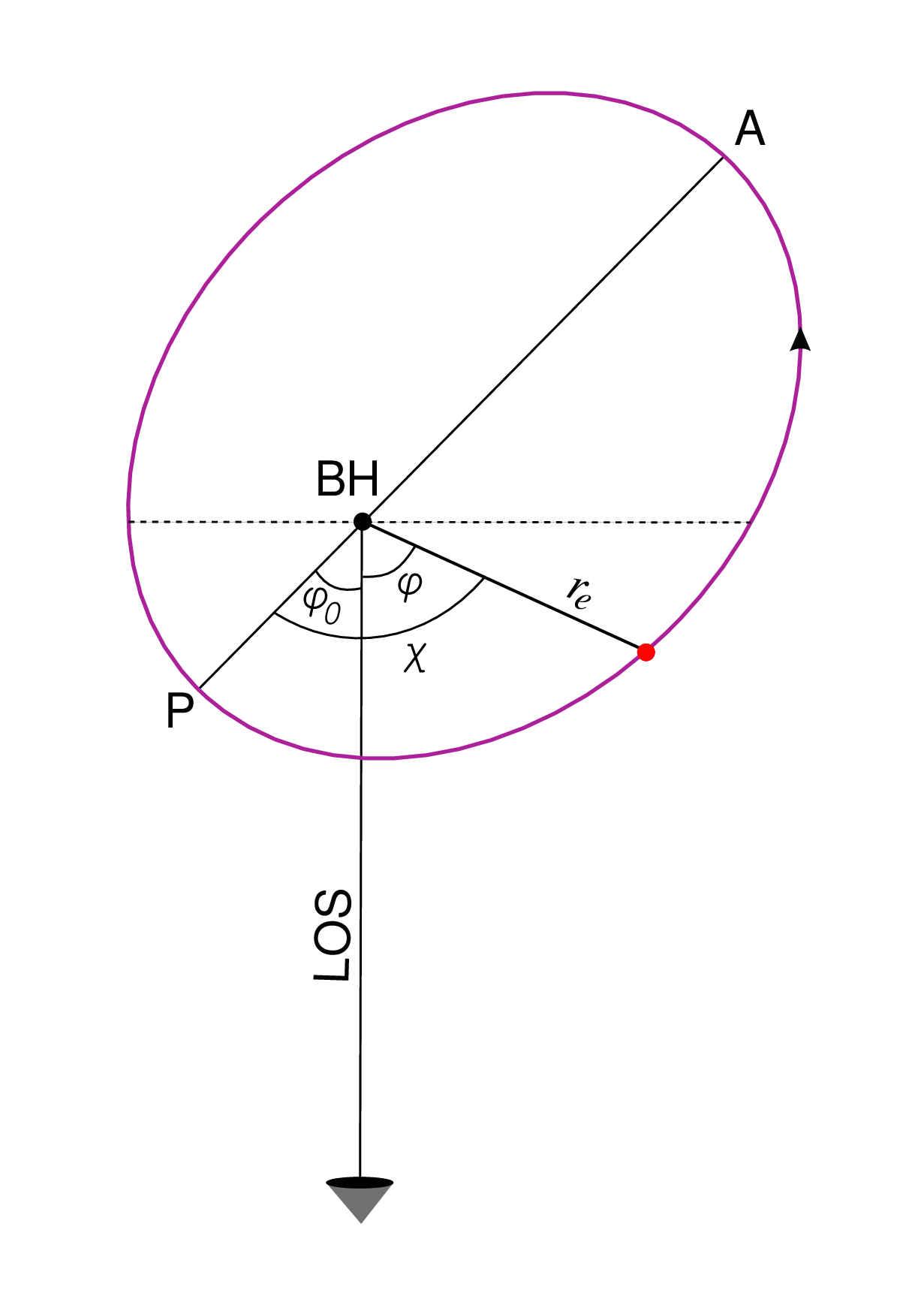}
\caption{Top left panel: The frequency shift $z_{Schw}$ versus the azimuthal angle in the Schwarzschild background for $M=1$, $p=100$, $e=0.3$, and $D=10^5$ considering the counterclockwise motion of the emitter. Different colors stand for various orientations of the orbits, i.e.,  $\varphi_0 = 0$ to $2\pi$ in intervals of $\pi/4$ (see bottom left panel). The deviation in the frequency shift due to orbital orientations is maximal close to the midline where $\varphi\approx\pm \pi /2$. Top right panel: Redshift rapidity versus the azimuthal angle for the same system. Bottom left panel: A schematic diagram showing the orientations of the elliptical orbits used to plot the top panels. Each color indicates an orbit with a fixed $\varphi_0$. The arrow denotes the counterclockwise motion of the emitter. Bottom right panel: The position of an emitter (red dot) is shown in an elliptical orbit along with the corresponding angular position $\varphi$ measured from the LOS, $\chi$ measured from the semi-major axis ($AP$), and $\varphi_0 = \chi - \phi$. $A$ and $P$ denote the apocenter and pericenter positions, respectively. The midline is shown by the horizontal dashed line. }
\label{z_vs_phi_elliptical_cc}
\end{figure*}

Now, let us discuss the relation between $\chi$ and $\varphi$. After taking derivative of Eq. (\ref{eqn:conicsection}) with respect to proper time of the emitter and doing some algebra using Eqs. (\ref{E_Lz}), (\ref{eqn:Lz}), and (\ref{eqn:ur}) we get
\begin{equation}
    \frac{d\varphi}{d\chi} = \left[ 1 - (3 + e \cos\chi) \frac{2M}{p}\right]^{-1/2}. \label{eqn:chi_varphi_relation}
\end{equation}
The above equation holds information about the general relativistic precession of the emitter's elliptical orbit around the black hole. To simplify our convention, we choose that $\chi$ and the azimuthal angle $\varphi$ increase in the same direction. We need to solve Eq. (\ref{eqn:chi_varphi_relation}) for $\chi$ to replace it in terms of $\varphi$ in our formalism. However, in the case of large $p$, we can get a simple linear relation, i.e., $\chi = \varphi + \varphi_0 $. The constant phase $\varphi_0$ is related to the orientation of the elliptical orbit with respect to the LOS (see Fig. \ref{z_vs_phi_elliptical_cc}).
 
In fact, the precession of the angle at which the radial coordinate attains its maximum/minimum value (apocenter/pericenter) is a property of the geodesic motion of the photon sources. 
The measurement of this quantity requires the observation of several orbital cycles of the emitter and hence, does not influence our
calculations of $M$ and $D$, 
which is based on measurements at specific points of the orbit of the emitter.

Furthermore, the light bending parameter is found similarly as before 
\begin{equation}
    b_\gamma = - \frac{r_e^{3/2} \sin(\varphi + \delta)}{\left[r_e - 2 M \sin^2(\varphi + \delta)\right]^{1/2}}. \label{eqn:b}
\end{equation}
Following the convention adopted in (\ref{lbp}), $b_\gamma$ is negative (positive) on the right (left) hand side of LOS. Therefore, 
for counterclockwise motion, $b_\gamma$ must have a minus sign on the right of LOS to yield a redshifted photon and vice versa for clockwise motion.

Now, we can use the definition of frequency shift in Eq. (\ref{freq_shift_circular}) for elliptical orbits and obtain
\begin{equation}
    1 + z_{Schw} = \frac{\omega_e}{\omega_d} = \frac{\left(E_\gamma U^t - g_{rr} k^r U^r -L_\gamma U^\varphi\right)\vert_e }{\left(E_\gamma U^t - g_{rr} k^r U^r -L_\gamma U^\varphi\right)\vert_d},
    \label{freq_shift_elliptical1}
\end{equation}
and as before, we consider the detector to be located far away from the emitter, thus obtaining
\begin{equation}
    1 + z_{Schw} = \left. \left( U^t - g_{rr} \frac{k^r}{E_\gamma} U^r - b_\gamma U^\varphi \right) \right\vert_e.
    \label{freq_shift_elliptical}
\end{equation}

Using Eqs. (\ref{rCompOf4momentum}) and (\ref{eqn:b}), we get 
\begin{equation}
\frac{k_e^r}{E_\gamma} = \frac{\cos(\varphi+\delta)}{\left[1-\frac{2M}{r_e}\sin^2(\varphi+\delta)\right]^{1/2}}. \label{eqn:krE}
\end{equation}

Here, $E_\gamma > 0$, and $k^r$ is positive if $-\pi/2 < (\varphi + \delta) < \pi/2 $ (emitter is located at the front), and negative if $ (\varphi + \delta) < -\pi/2 $ and  $ \varphi + \delta > \pi/2 $ (emitter is located at the back). Thus, the sign of $k^r/E_\gamma$ is automatically defined by $(\varphi + \delta)$.

Now, replacing Eqs. (\ref{E_Lz}), (\ref{eqn:ur}), (\ref{eqn:b}), and (\ref{eqn:krE}) in Eq. (\ref{freq_shift_elliptical}), we can obtain an exact formula for the frequency shift including arbitrary values of the eccentricity $e$ as below
\begin{eqnarray}
1+z_{Schw} &=&\frac{1}{\sqrt{c_{1}}fr_{e}}\left( \sqrt{c_{1}}c_{3}r_{e}+%
\frac{c_{2}fMr_{e}\sin (\delta +\varphi )}{c_{4}^{2}r_{e}}\right.   \nonumber
\\
&&\left. -c_{2}er_{e}\sqrt{f-4c_{4}^{2}}\sin \chi \cos (\delta +\varphi
)\right) ,
\label{RedNonFinal}
\end{eqnarray}
where the parameters $f$\ and $c_{i}$ ($i=1,2,3,4$) are given by%
\begin{equation}
f=1-\frac{2M}{r_{e}},\ \ \ c_{1}=1-\frac{2M\sin ^{2}(\delta +\varphi )}{r_{e}%
},
\end{equation}%
\begin{equation}
c_{2}=\sqrt{\frac{M}{r_{e}(1+e\cos \chi )-\left( e^{2}+3\right) M}},
\end{equation}%
\begin{equation}
c_{3}=\frac{c_{2}}{c_{4}}\sqrt{4c_{4}^{4}\left( 1-e^{2}\right) -4c_{4}^{2}+1}%
,
\end{equation}%
\begin{equation}
c_{4}=\sqrt{\frac{M}{r_{+}(1+e\cos \chi )}}.
\end{equation}%

Now, in what follows we consider a first-order approximation in $e$. Thus, in the far-away observer limit $D \gg r_e$, we get
\begin{align}
1 & +z_{Schw} = \sqrt{\frac{r_{e}}{r_{e}-3M}} \Biggl[ 1 \pm \mathcal{A} \sin(\varphi + \delta) \nonumber \\ 
& + \Biggl( \frac{M(r_e-6M)}{2(r_e-3M)(r_e-2M)} \cos\chi \nonumber \\ 
& \pm \frac{\mathcal{A}(r_e-6M)}{2(r_e-3M)} \cos\chi \sin(\varphi + \delta) \nonumber \\ 
& \mp \frac{\mathcal{A}\sqrt{r_e(r_e-6M)}}{r_e-2M} \cos(\varphi + \delta) \sin\chi \Biggl) e \Biggl] +  ~\mathcal{O}(e^2),
\label{zSchw_elliptical}
\end{align} 
where $\chi \approx \varphi + \varphi_0$ and
\begin{equation}
    \mathcal{A} = \sqrt{\frac{M}{r_{e}-2M\sin ^{2}\left( \varphi +\delta \right) }}.\label{eqn:Afactor}
\end{equation}

In the above Eq. (\ref{zSchw_elliptical}), the upper (lower) sign stands for the counterclockwise (clockwise) motion of the emitter. Note that a redshifted photon is generated in two ways: on the right-hand side of the LOS for a counterclockwise motion and on the left of LOS for a clockwise motion. Both ways are physically equivalent which can be confirmed from Eq. (\ref{zSchw_elliptical}). 

In the Newtonian limit, i.e., $D\gg r_e\gg M$, we can further simplify Eq. (\ref{zSchw_elliptical}) as
\begin{align}
& z_{N} =\pm \sqrt{\frac{M}{r_e}} \sin\left( \varphi +\delta \right) \times \nonumber\\
& \Biggl[ 1 + \Biggl( \frac{1}{2}\cos\chi -\sin\chi \cot\left( \varphi +\delta \right)  \Biggl) e \Biggl]
+ \mathcal{O}\left(\frac{M}{r_e}\right).
\end{align}

\section{Redshift rapidity in case of elliptical orbit of the emitter}

Here, we calculate the redshift rapidity in the case of an eccentric emitter orbit. Next, we shall present formulas for small eccentricity as well. We define redshift rapidity as 
\begin{align}
    \frac{dz}{dt} &= \frac{dz}{d\tau}\frac{d\tau}{dt} = \frac{1}{U_e^t} \Biggl[ \frac{d U_e^t}{d\tau} - b_\gamma \frac{d U_e^\phi}{d\tau} - \frac{db_\gamma}{d\tau} U_e^\phi  \nonumber \\
    & - g_{rr} \frac{d U_e^r}{d\tau}\frac{k_e^r}{E_\gamma} - \frac{d g_{rr}}{d\tau}U_e^r \frac{k_e^r}{E_\gamma} - g_{rr} U_e^r \frac{d}{d\tau}\left(\frac{k_e^r}{E_\gamma}\right) \Biggl].  \label{dzdt_a}
\end{align}
By using Eqs. (\ref{4vel_norm1})-(\ref{E_Lz}), we get
\begin{align}
    \frac{dU_e^r}{d\tau} &= \frac{1}{r_e}\left[\frac{L_z^2}{r_e^2}\left(1-\frac{3M}{r_e}\right)-\frac{M}{r_e}\right],\\
    \frac{dU_e^\phi}{d\tau} &= -\frac{2L_z}{r_e^3} U_e^r,\\
    \frac{dU_e^t}{d\tau} &= -\frac{2 M E}{\left(r_e-2M\right)^2} U_e^r.
\end{align}

Again, from the null condition $k_\mu k^\mu = 0$, we can obtain the derivative of $k_e^r/E_\gamma$ as follows
\begin{align}
    \frac{d}{d\tau}\left(\frac{k_e^r}{E_\gamma}\right) = \frac{ (r_e-3 M)b_\gamma^2 U_e^r - r_e (r_e-2M)b_\gamma \frac{db_\gamma}{d\tau}}{r_e^{5/2} \sqrt{b_\gamma^2 (2 M-r_e)+r_e^3}}.
\end{align}

On the other hand, by employing the chain rule, we have
\begin{equation}
\frac{db_\gamma}{d\tau} = \left( \frac{\partial b_{\gamma }}{\partial r } U_e^r + \frac{\partial b_{\gamma }}{\partial \varphi } U_e^\varphi +\frac{%
\partial b_{\gamma }}{\partial \delta }\frac{\partial \delta }{\partial
\varphi } U_e^ \varphi \right),
\end{equation}%
where $\frac{\partial\delta}{\partial\varphi}$ is obtained from Eq. (\ref{deltaVSphi}). By calculating the terms presented in Eq. (\ref{dzdt_a}), we can find the redshift rapidity for arbitrary eccentric orbits as follows
\begin{widetext}
\begin{eqnarray}
\frac{dz_{Schw}}{dt} &=&-\frac{f}{c_{3}r_{e}^{2}}\left[ \frac{c_{2}c_{8}M}{%
c_{4}^{2}}+\frac{2c_{3}c_{4}c_{5}eM\sin \chi }{f^{2}}\right. -\frac{M\cos
(\delta +\varphi )}{\sqrt{c_{1}}f}\left( 1+\frac{c_{2}^{2}M(3M-r_{e})}{%
c_{4}^{2}r_{e}^{2}}\right) -\frac{2c_{4}^{2}c_{5}^{2}e^{2}M\sin ^{2}\chi
\cos (\delta +\varphi )}{\sqrt{c_{1}}f^{2}}  \nonumber \\
&&+\frac{2c_{4}c_{5}e\sin \chi \sin (\delta +\varphi )}{\sqrt{c_{1}}}\left.
\left( \frac{c_{2}M}{c_{4}^{2}}+\frac{\sqrt{c_{1}}c_{8}r_{e}}{2\sqrt{%
c_{1}-f\sin ^{2}(\delta +\varphi )}}-\frac{c_{4}c_{5}e(3M-r_{e})\sin \chi
\sin (\delta +\varphi )}{2f\sqrt{c_{1}-f\sin ^{2}(\delta +\varphi )}}\right) %
\right] , \label{RRGe}
\end{eqnarray}
\end{widetext}
where the parameters $c_{i}$ ($i=5,6,7,8$) are given by
\begin{equation}
c_{5}=\sqrt{\frac{r_{e}(1+e\cos \chi )-2M(3+e\cos \chi )}{r_{e}(1+e\cos \chi
)-\left( e^{2}+3\right) M}},
\end{equation}%
\begin{equation}
c_{6}=-\frac{\cos (\delta +\varphi )}{\sqrt{c_{1}}}\left( \frac{2M\sin
^{2}(\delta +\varphi )}{c_{1}}+r_{e}\right) ,
\end{equation}%
\begin{equation}
c_{7}=\frac{\sin (\delta +\varphi )}{\sqrt{c_{1}}}\left( \frac{M\sin
^{2}(\delta +\varphi )}{c_{1}r_{e}}-1\right) ,
\end{equation}%
\begin{eqnarray}
c_{8} &=&\frac{c_{2}c_{6}(1+e\cos \chi )(D\cos \varphi -r_{e})}{%
D^{2}+r_{e}^{2}-2r_{e}D\cos \varphi }  \nonumber \\
&&+\frac{c_{2}c_{6}M}{c_{4}^{2}r_{e}^{2}}+c_{4}c_{5}c_{7}e\sin \chi .
\end{eqnarray}

This redshift rapidity formula (\ref{RRGe}) takes the following form in the limit of a faraway observer, $D\gg r_e$, as
\begin{align}
    \frac{dz}{dt} &= \frac{\mathcal{A}^3 M^{-1/2} }{\sqrt{r_e-3M} }\cos(\varphi + \delta) ~+ \nonumber\\
    &\sqrt{\frac{r_e-6M}{r_e-3M}} \times \Biggl[ \mp \frac{2M^{3/2}}{r_e^{3/2}(r_e-2M)} \sin\chi  \nonumber \\
    & + \mathcal{A}^3 M^{1/2}\Biggl( \frac{\sqrt{r_e-6M}}{2(r_e-2M)(r_e-3M)} \cos(\varphi + \delta) \cos\chi \nonumber\\
    & + \frac{2 \sqrt{r_e-6M} } {r_e(r_e-2M)} \sin^2(\varphi + \delta) \cos(\varphi + \delta) \cos\chi \nonumber\\
    & + \frac{1}{ r_e^{3/2} } \sin^3(\varphi + \delta) \sin\chi  \Biggl) \Biggl] e + \mathcal{O}(e^2),
    \label{dzdt_b}
\end{align}
where $\mathcal{A}$ is defined in Eq. (\ref{eqn:Afactor}). For the counterclockwise motion of the emitter, redshift rapidity is plotted for various orientations of elliptical orbits (see top right panel of Fig. \ref{z_vs_phi_elliptical_cc}). Similar to the frequency shift, we see from Eq. (\ref{dzdt_b}) that redshift rapidity on the right-hand side of LOS for counterclockwise motion (upper sign) is equivalent to that on the left-hand side of LOS for clockwise motion (lower sign). 

In the Newtonian limit $D\gg r_e\gg M$, Eq. (\ref{dzdt_b}) is simplified further as follows
\begin{align}
    \frac{dz_N}{dt} &= \frac{M }{r_e^2}\cos (\varphi + \delta) \mp e \frac{2 M^{3/2}}{r_e^{5/2}}\sin\chi +~\mathcal{O}\left(\frac{M^2 }{r_e^3}e \right).
    \label{dzdt_N}
\end{align}

It is worth mentioning that there exist measurements of the redshift rapidity for some
megamaser systems in the accretion disks of supermassive black holes hosted
at the center of AGNs, such as UGC 3789 \cite{MCP2}, NGC 6264 \cite{MCPV}, NGC 5765b \cite{MCPVIII}, CGCG
074-064 \cite{MCPXI} that are reported as \textquotedblleft
accelerations" of the photon sources. These measured redshift rapidities range $[-1.68,3.48]~km/s/yr$ for test particles emitting at the
midline and lie within $[0.9,8.4]\ km/s/yr$ for photon sources located by the
observer's line of sight. However, the corresponding errors lie in the
ranges $[0.50,3.36]$ and $[0.20,3.07]\ km/s/yr$, which are quite big
compared to the measured quantities.

Finally, one should note that, unlike the circular case, we cannot obtain analytic formulas for $M$ and $D$ in terms of observational quantities due to the complexity of the expressions (\ref{RedNonFinal}) and (\ref{RRGe}). 
In this case, equations (42) and (56) can be inverted numerically or one needs to rely on a Bayesian statistical fit of observational data and use Eqs. (\ref{RedNonFinal}) and (\ref{RRGe}) simultaneously to determine $M$ and $D$ for non-circular orbits.

\section{Error estimation of the frequency shift and the redshift rapidity formula}

In Eq. (\ref{zSchw_elliptical}) and Eq. (\ref{dzdt_b}), we have expanded the frequency shift and the redshift rapidity up to the first order in eccentricity. Naturally, it is necessary to discuss the error included while ignoring the higher-order terms. To quantify that, we define the relative error in frequency shift as follows 
\begin{equation}
    \frac{\delta z}{1+z} = \left\vert \frac{(1+z) - (1+z_{Schw})}{1+z} \right\vert,
    \label{eqn:error_z}
\end{equation}
where $1+z$ is the exact value calculated numerically using Eq. (\ref{freq_shift_elliptical}) and $1+z_{Schw}$ is obtained from the approximate expression Eq. (\ref{zSchw_elliptical}). Similarly, the relative error in the redshift rapidity is defined as
\begin{equation}
    \frac{\delta \Dot{z}}{\Dot{z}} = \left\vert \frac{\Dot{z} - \Dot{z}_{Schw}}{\Dot{z}\vert_{\varphi=0}} \right\vert.
    \label{eqn:error_dzdt}
\end{equation}

\begin{figure*}[ht]
\centering
\includegraphics[scale=.4]{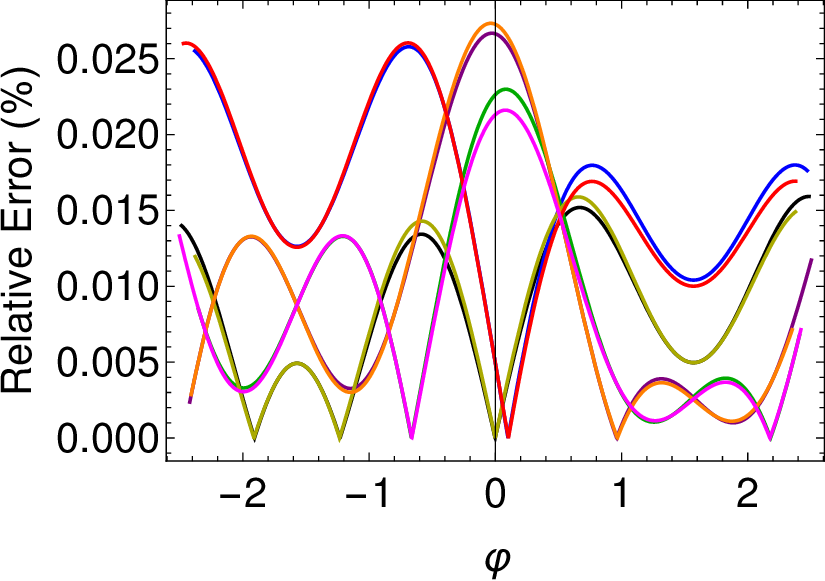}
\includegraphics[scale=.4]{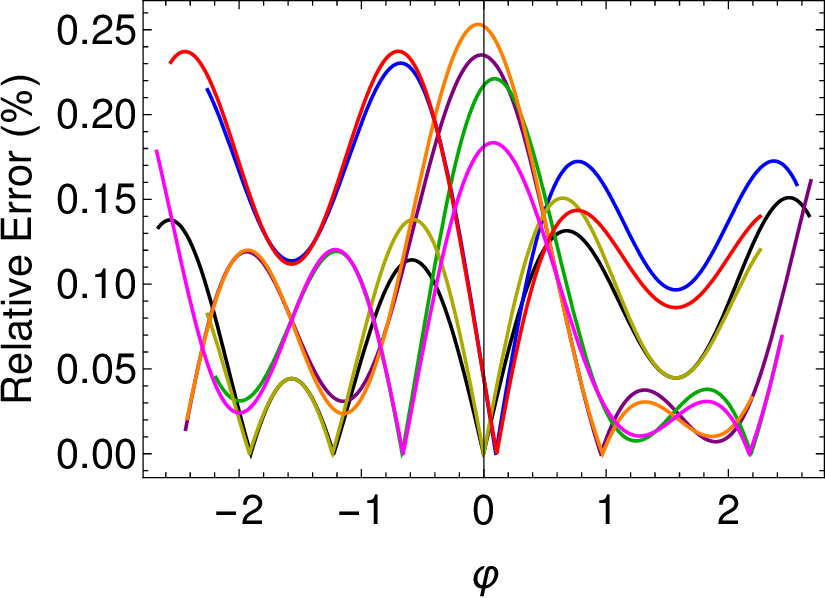}
\includegraphics[scale=.4]{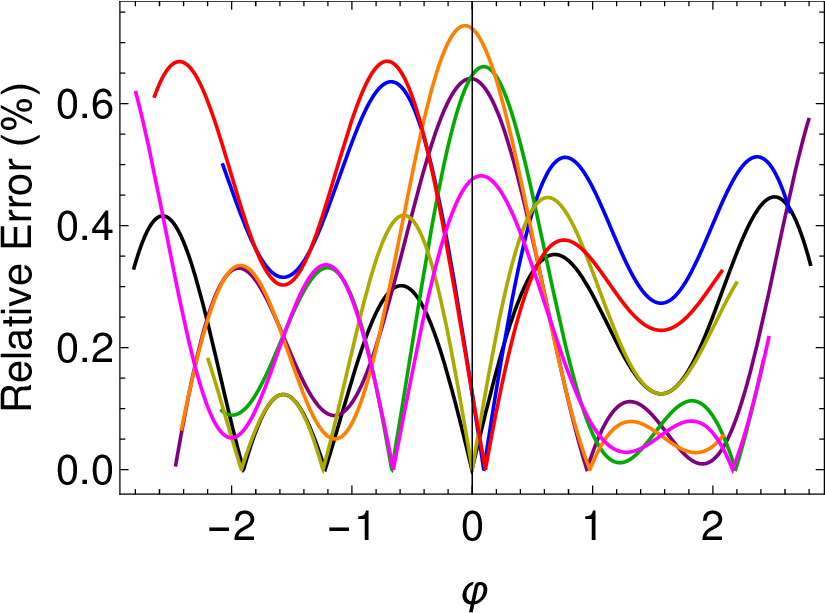}
\includegraphics[scale=.4]{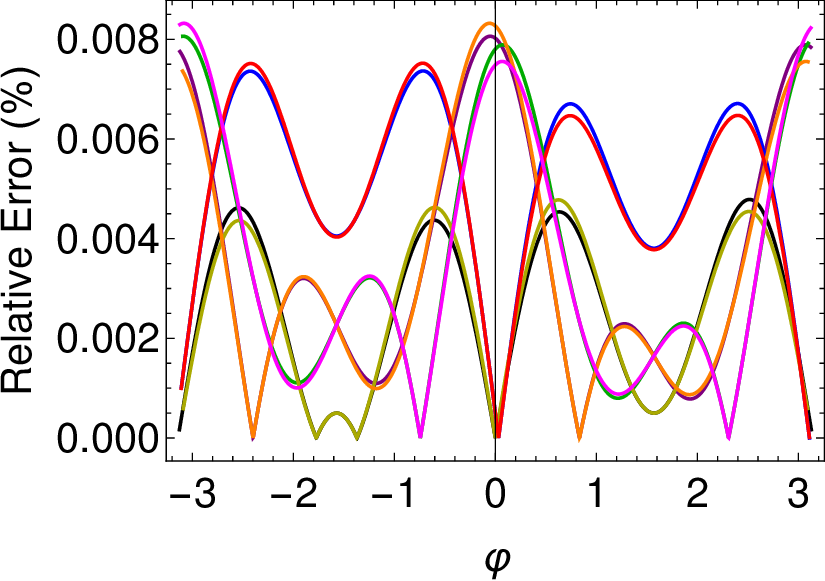}
\includegraphics[scale=.4]{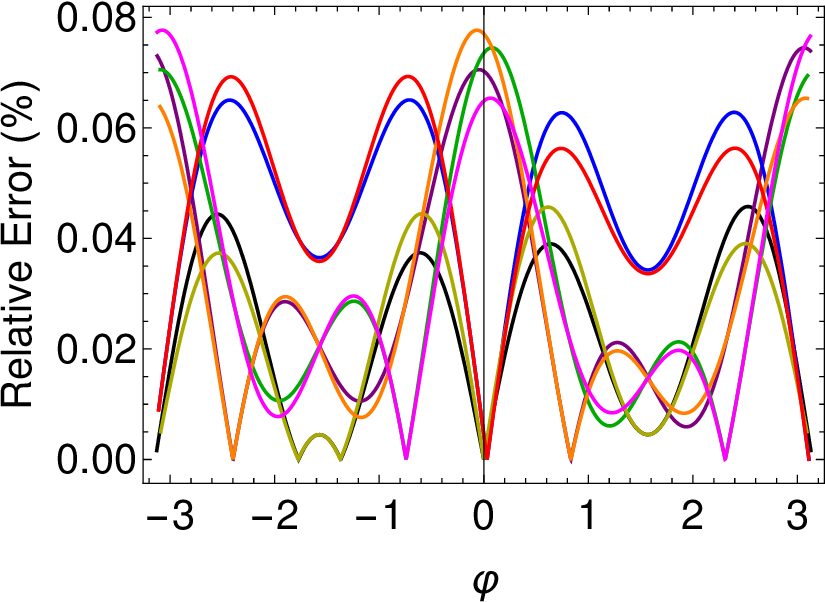}
\includegraphics[scale=.4]{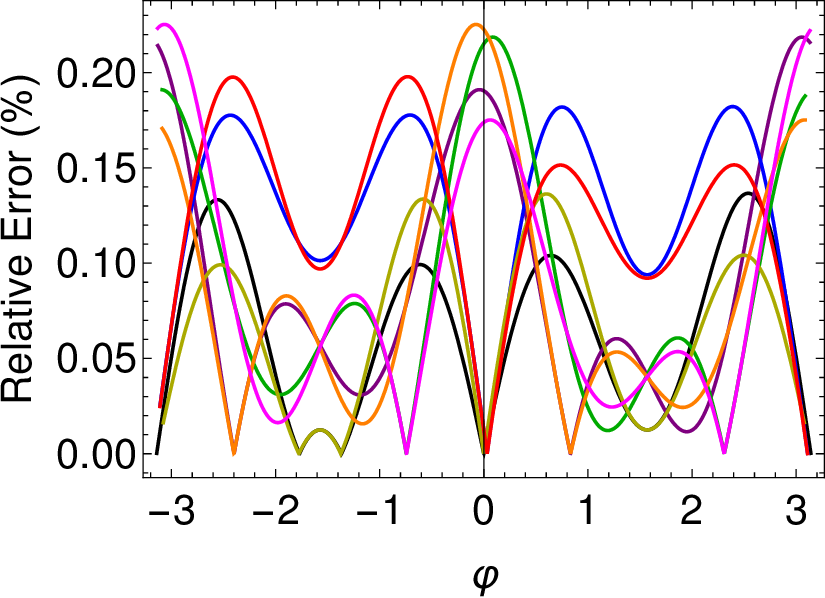}

\caption{Relative percentage error of frequency shift given by Eq. (\ref{eqn:error_z}) for various values of orbital parameters $p$ and $e$. The upper (lower) panels refer to $p = 100$ ($p=1000$). Columns from left to right indicate $e= 0.1, 0.3, 0.5$, respectively. We have set $M=1$, and a large value of $D=1000p$. For various colors in each plot, see the bottom panel of Fig. \ref{z_vs_phi_elliptical_cc}. }
\label{error_z}
\end{figure*}

\begin{figure*}[ht]
\centering
\includegraphics[scale=.4]{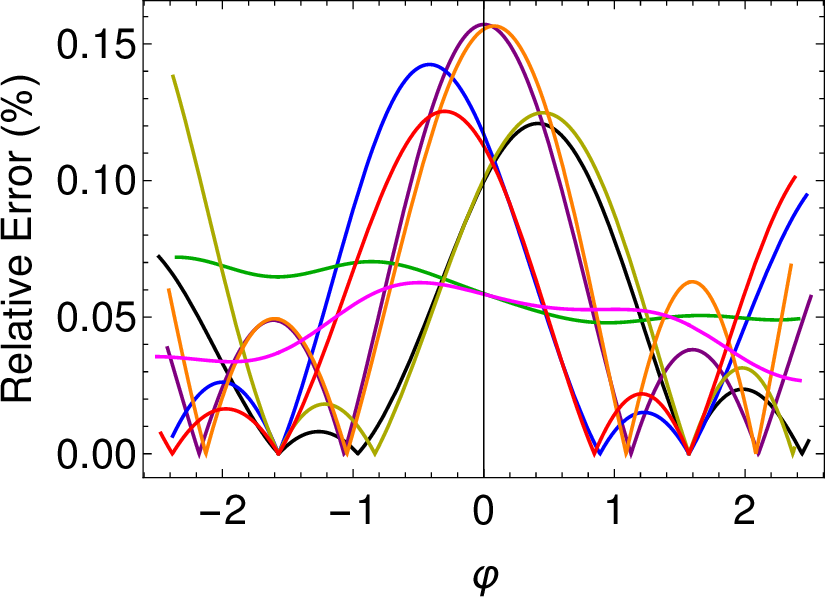}
\includegraphics[scale=.4]{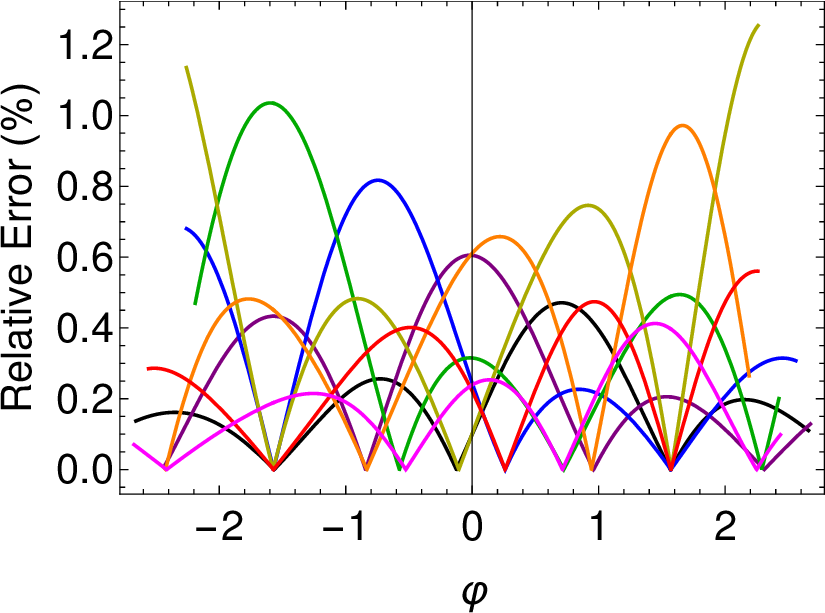}
\includegraphics[scale=.4]{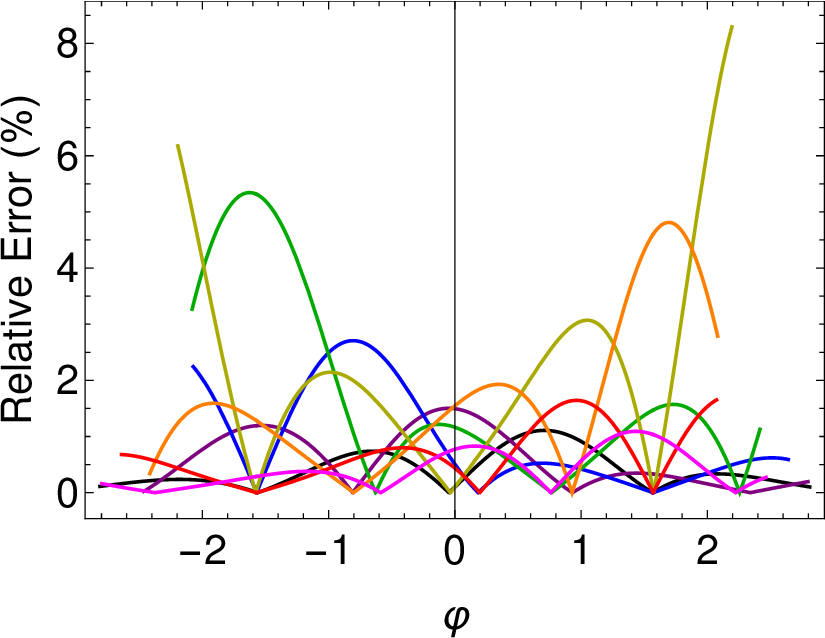}
\includegraphics[scale=.4]{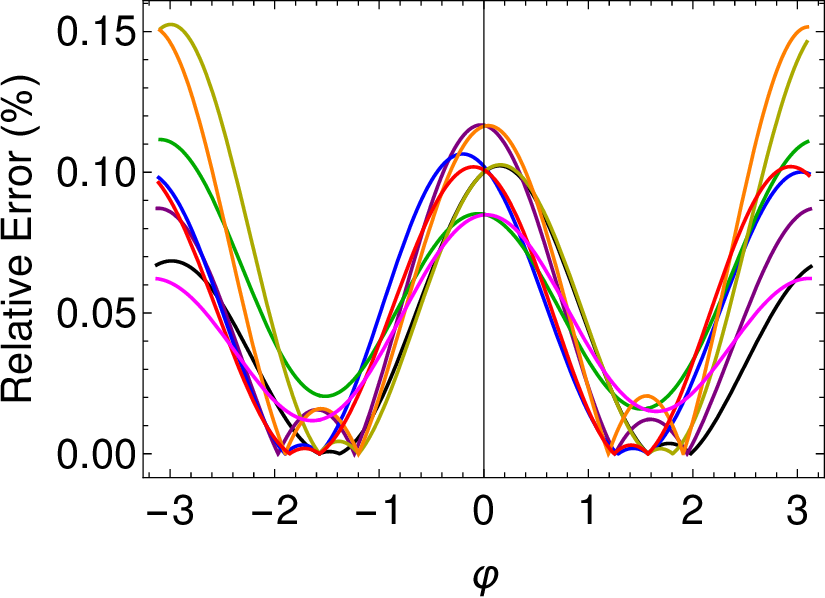}
\includegraphics[scale=.4]{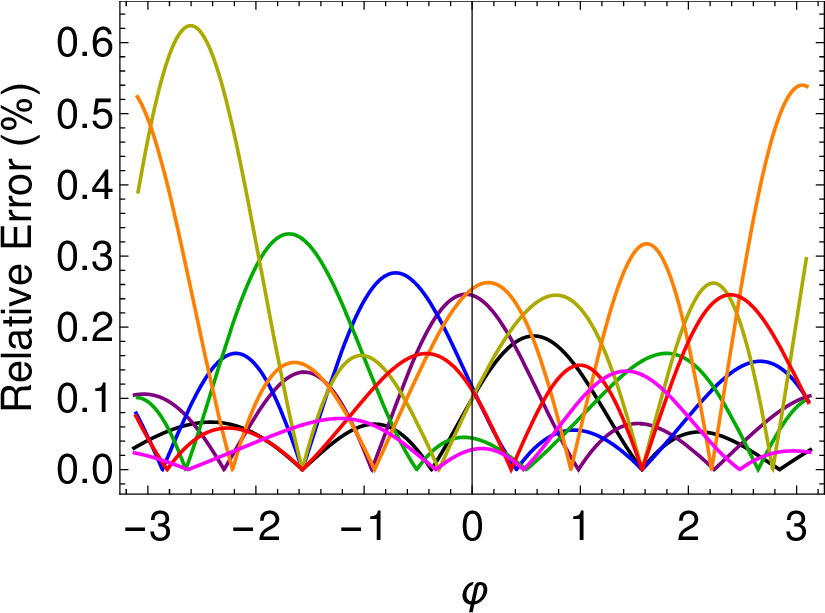}
\includegraphics[scale=.4]{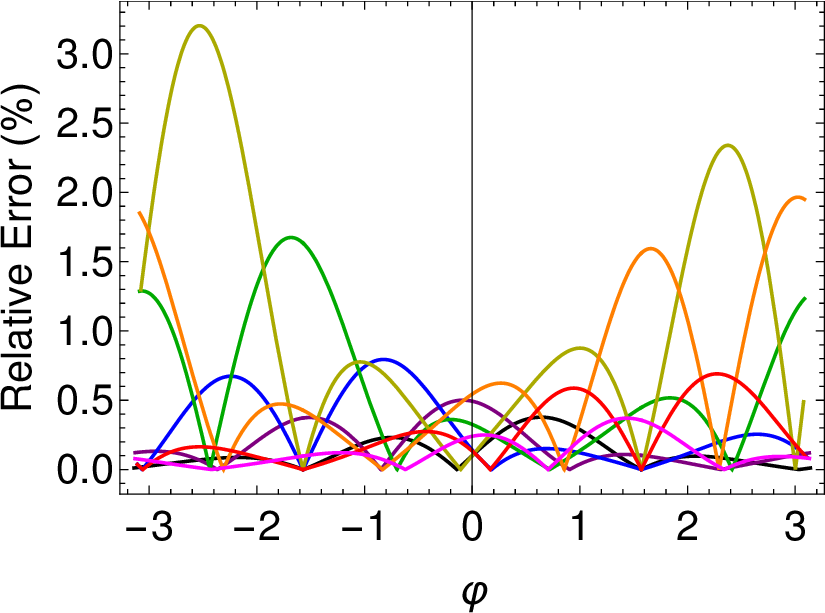}

\caption{Relative percentage error of redshift rapidity given by Eq. (\ref{eqn:error_dzdt}) for various values of orbital parameters $p$ and $e$. The upper (lower) panels refer to $p = 100$ ($p=1000$). Columns from left to right indicate $e= 0.1, 0.3,$ and $0.5$, respectively. $M=1$, and $D=1000p$. For various colors in each plot, see the bottom panel of Fig. \ref{z_vs_phi_elliptical_cc}.}
\label{error_dzdt}
\end{figure*}

Here, $\Dot{z}$ is the exact value of the redshift rapidity calculated numerically using Eq. (\ref{dzdt_a}), and $\Dot{z}_{Schw}$ is defined by our approximate expression Eq. (\ref{dzdt_b}). The denominator is calculated at $\varphi=0$ to avoid the divergence near $\varphi = \pi/2$ where $\Dot{z}$ becomes zero (see Fig. \ref{z_vs_phi_elliptical_cc}). In Figs. \ref{error_z} and \ref{error_dzdt}, we have shown the relative percentage error against $\varphi$ for various emitter orbits. As expected, the error is higher for higher $e$ and lower $p$. As seen from the figures, the errors in frequency shift near the midline ($\varphi \approx \pm \pi/2$) are within $0.1\%$ for $e=0.1$, and $1\%$ for $e=0.3$, whereas it is relatively significant for $e=0.5$ or above. Similarly, for redshift rapidity, the error remains within an acceptable range for eccentricity $e < 0.3$. 

\section{Discussion and final remarks}

In this paper, we have presented two relations for the mass-to-distance
ratio of the Schwarzschild black hole in terms of observable frequency shifts
for three spacial points of the circular motion of orbiting massive
particles: on the midline and close to the LOS. We further calculated the derivative of the frequency shift in the Schwarzschild background with
respect to proper time in order to introduce the redshift rapidity. Then, we
employed the redshift rapidity to express the Schwarzschild black hole
mass and its distance from the Earth just in terms of observational
quantities. In this formalism, the redshift rapidity is also a general relativistic invariant observable element, indicating the proper time evolution of the frequency shift in the
Schwarzschild spacetime.

We have extracted concise and elegant analytic formulas that allow us
to disentangle mass and distance to the black hole in the Schwarzschild background and
compute these parameters separately, not the mass-to-distance ratio that we
have estimated before in  \cite{ApJL,TXS,TenAGNs,FiveAGNs}. Our analytic
formulas have been obtained on the midline and close to the LOS
whereas the general relations could be employed in black hole parameter estimation studies. 

We also computed the redshift and redshift rapidity for eccentric orbits of the emitter and calculated their relative error with respect to the numerical exact value of these quantities given by Eqs. (\ref{freq_shift_elliptical}) and (\ref{dzdt_a}), respectively. A simple analysis shows that the errors for these quantities at first order in $e$ are within 1\% for $e=0.3$.

The next step in this research direction would be estimating the
mass of supermassive Schwarzschild black holes hosted at the core of
AGNs and their distance to the Earth with the help of the general
relativistic formalism developed in the present paper. To do so, we need to perform a Bayesian statistical fit with the help of Eqs. (\ref{RedNonFinal}) and (\ref{RRGe}) as well as precise measurements of the positions of the photon sources and their redshift and redshift rapidities. A task that we shall address in future works.

\section*{Acknowledgements}

We thank D. Villaraos and A. Gonz\'alez-Ju\'arez for fruitful discussions.
All authors are grateful to CONACYT for support under Grant No.
CF-MG-2558591; M.M. also acknowledges SNI and was supported by CONACYT
through the postdoctoral Grant No. 31155. 
P.B. and A.H.-A. acknowledge financial support from the Science and Engineering Research Board, Government of India, File Number PDF/2022/000332.
A.H.-A. and U.N. thank SNI and
PRODEP-SEP and were supported by Grants VIEP-BUAP and CIC-UMSNH, respectively. U.N. also acknowledges support under Grant No. CF-140630. 

\appendix*

\section{($\protect\varphi +\protect\delta $)-dependent light bending
parameter and the angular distance $\protect\delta $ as a function of the
azimuthal angle $\protect\varphi $}

\label{bphi}

\begin{figure*}[t]
\centering
\includegraphics[scale=.9]{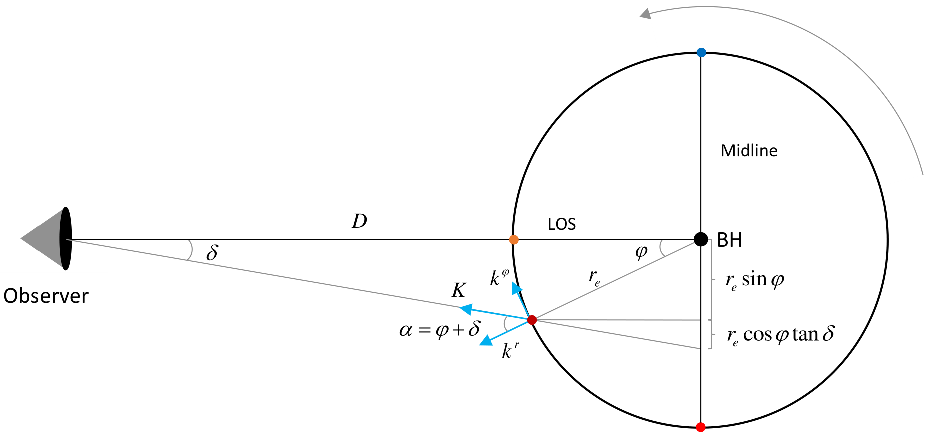}
\caption{The geometrical illustration of the bidimensional vector $K$, and
the relation between the azimuthal angle $\protect\varphi $ and the angular
distance $\protect\delta $.}
\label{PhiDeltaFig}
\end{figure*}
\begin{figure}[t]
\centering
\includegraphics[scale=.9]{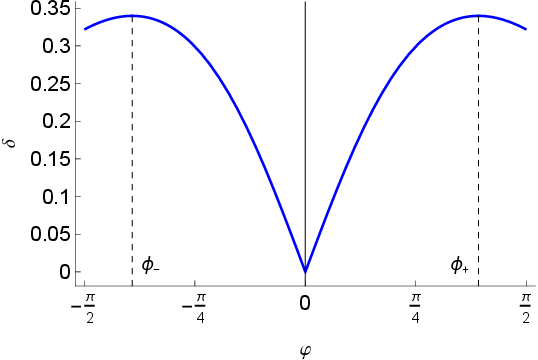}
\caption{The profile of the angular distance versus the azimuthal angle for $D=3r_{e}$. The vertical dashed lines denote $\phi _{\pm}=\pm \arccos(1/3)$ where $d\delta /d\varphi $\ vanishes.}
\label{delatVSphiPlot}
\end{figure}

In a similar manner to \cite{PRDbhmn}, we obtain the light bending parameter
for a general point of the circular orbit in the equatorial plane. However,
unlike the previous work\ where the angular distance $\delta $\ was
neglected due to the large distances between the black holes and the Earth,
we consider the contribution of this measurable parameter in our analysis.
This is because even though the angular distance is very small, its
variation with respect to time, especially close to the LOS, is
quite important. The other aim of this Appendix is to show the mathematical
relation between the observable angular distance $\delta $ and the
unobservable azimuthal angle $\varphi $ if the emitter particle is far enough from the black hole.

In order to obtain the light bending parameter of photons coming from a
general point of the circular orbit in the equatorial plane (see Fig. \ref%
{PhiDeltaFig}), we take into account the equation of motion of photons in
the equatorial plane ($k^{\theta }=0$) of the Schwarzschild spacetime (\ref%
{metric}) 
\begin{equation}
g_{\mu \nu }k^{\mu }k^{\nu }=g_{tt}\left( k^{t}\right) ^{2}+g_{rr}\left(
k^{r}\right) ^{2}+g_{\varphi \varphi }\left( k^{\varphi }\right) ^{2}=0,
\label{PhotonEOM}
\end{equation}%
where $k^{t}$\ and $k^{\varphi }$\ can be found through the temporal Killing
vector field $\xi ^{\mu }=\delta _{0}^{\mu }$\ and the rotational Killing
vector field $\psi ^{\mu }=\delta _{3}^{\mu }$ as follows 
\begin{equation}
E_{\gamma }=-\xi _{\mu }k^{\mu }=-g_{tt}k^{t},  \label{tCompOf4momentum}
\end{equation}%
\begin{equation}
L_{\gamma }=\psi _{\mu }k^{\mu }=g_{\varphi \varphi }k^{\varphi },
\label{PhiCompOf4momentum}
\end{equation}%
which $E_{\gamma }$ and $L_{\gamma }$ are the energy and angular momentum of
the photons, respectively.

By introducing (\ref{tCompOf4momentum})\ and (\ref{PhiCompOf4momentum}) into
(\ref{PhotonEOM}), the equation of motion\ takes the following form%
\begin{equation}
g_{rr}\left( k^{r}\right) ^{2}+\frac{E_{\gamma }^{2}}{g_{tt}}+\frac{%
L_{\gamma }^{2}}{g_{\varphi \varphi }}=0,
\end{equation}%
that can be employed to express $k^{r}$ in terms of the constants of motion
and metric components as below\ 
\begin{equation}
\left( k^{r}\right) ^{2}=-\frac{1}{g_{rr}}\left( \frac{E_{\gamma }^{2}}{%
g_{tt}}+\frac{L_{\gamma }^{2}}{g_{\varphi \varphi }}\right) .
\label{rCompOf4momentum}
\end{equation}

Now, by considering the non-vanishing angular distance $\delta \neq 0$, we
geometrically introduce the auxiliary bidimensional vector $K$ defined by
the following decomposition (see Fig. \ref{PhiDeltaFig} for more details)%
\begin{equation}
k^{r}=K\cos \alpha ,  \label{kr}
\end{equation}%
\begin{equation}
r_{e}k^{\varphi }=K\sin \alpha ,  \label{kphi}
\end{equation}%
with $\alpha$ the angle between the photon ray direction and its radial component at the emitter position and%
\begin{equation}
K^{2}=\left( k^{r}\right) ^{2}+r_{e}^{2}\left( k^{\varphi }\right) ^{2}.
\label{Kdef}
\end{equation}

Hence, by substituting Eqs. (\ref{PhiCompOf4momentum})\ and (\ref%
{rCompOf4momentum})\ into (\ref{Kdef}), one can find $K^{2}$ versus the
constants of motion and metric components as follows
\begin{equation}
K^{2}=-\frac{1}{g_{rr}}\left( \frac{E_{\gamma }^{2}}{g_{tt}}+\frac{L_{\gamma
}^{2}}{g_{\varphi \varphi }}\right) +r_{e}^{2}\frac{L_{\gamma }^{2}}{%
g_{\varphi \varphi }^{2}}.
\end{equation}

On the other hand, by introducing Eq. (\ref{kr})\ into (\ref{rCompOf4momentum}%
), we can find a similar relation for $K^{2}$ as%
\begin{equation}
K^{2}=-\frac{1}{g_{rr}\cos ^{2}\alpha }\left( \frac{%
E_{\gamma }^{2}}{g_{tt}}+\frac{L_{\gamma }^{2}}{g_{\varphi \varphi }}\right)
.
\end{equation}

Equating previous relations gives an equation for the $\alpha $-dependent light bending parameter $b_{\gamma }$ as below%
\begin{equation}
\left( g_{tt}b_{\gamma }^{2}+g_{\varphi \varphi }\right) g_{\varphi \varphi
}\tan ^{2}\alpha +r_{e}^{2}g_{rr}g_{tt}b_{\gamma
}^{2}=0,
\end{equation}%
that leads to the following solution for $b_{\gamma }$ 
\begin{equation}
b_{\gamma }=\frac{-g_{\varphi \varphi }\sin \alpha }{%
\sqrt{-g_{tt}}\sqrt{g_{\varphi \varphi }\sin ^{2}\alpha +r_{e}^{2}g_{rr}\cos ^{2}\alpha }},
\label{bGamma}
\end{equation}%
which is the light bending parameter for an arbitrary point of the circular
orbit on the equatorial plane.

On the other hand, if we assume the emitter is far enough from the black hole
and consider the geometrical configuration illustrated in Fig. \ref%
{PhiDeltaFig} we can write $\alpha\approx \varphi + \delta$, where, as mentioned earlier, $\delta$ is the is the observable angular distance and $\varphi$ is the unobservable azimuthal angle. 

In this case, we can obtain a relation between  $%
\delta $ and  $\varphi $. This is necessary
to express $\varphi $\ in terms of the observable quantity $\delta $ in
order to obtain $M$\ and $D$\ just in terms of observational elements, hence
being able to break the degeneracy of the $M/D$\ ratio.

By taking into account the right triangle in Fig. \ref{PhiDeltaFig}, one can
immediately identify the following relation between the set $\left\{
D,r_{e};\varphi ,\delta \right\} $ 
\begin{equation}
D\tan \delta =r_{e}\left( \sin \varphi +\cos \varphi \tan \delta \right) ,
\end{equation}%
where only $\delta $\ is observable and the rest of the parameters are
unknown in the case of black holes since we cannot identify the black hole
position by observations. After doing some straightforward simplifications,
this relation reduces to 
\begin{equation}
D\sin \delta =r_{e}\sin \left( \varphi +\delta \right) ,  \label{PhiDeltaGR}
\end{equation}%
in which for the far away detectors where $D\gg r_{e}$\ and $\delta
\rightarrow 0$, we have%
\begin{equation}
r_{e}\approx D\delta _{m}\left( 1+\frac{\delta _{m}^{2}}{3}\right) ,\text{ \
for \ }\varphi =\pm \frac{\pi }{2},  \label{AppMid}
\end{equation}%
\begin{equation}
r_{e}\approx \frac{D\delta _{s}}{\varphi _{s}},\text{ \ for \ }\varphi
\rightarrow 0,  \label{AppLOS}
\end{equation}%
for the midline $\varphi =\pm \pi /2$\ and close to the LOS $%
\varphi \rightarrow 0$, respectively.

Now, we solve the relation (\ref{PhiDeltaGR})\ in order to express $\delta $%
\ in terms of the rest of the parameters. This equality has four solutions, and
we choose the physical one as below 
\begin{equation}
\delta =\arccos \left( \frac{D-r_{e}\cos \varphi }{\sqrt{%
D^{2}+r_{e}^{2}-2Dr_{e}\cos \varphi }}\right) ,  \label{deltaVSphi}
\end{equation}%
since as $\varphi $\ decreases/increases, $\delta $\ should also
decrease/increase in the same direction (see Fig. \ref{PhiDeltaFig}).

Fig. \ref{delatVSphiPlot} illustrates the behavior of $\delta $ versus  $\varphi $. As one can see from this figure, if we solve the azimuthal angle $\varphi $%
\ in terms of $\delta $, we shall have four expressions as follows%
\begin{equation}
\varphi =\left\{ 
\begin{array}{cc}
\pm \delta \mp \arcsin \left( \frac{D\sin \delta }{r_{e}}\right) , & \phi
_{-}\leq \varphi \leq \phi _{+} \\ 
\pm \pi \mp \delta \mp \arcsin \left( \frac{D\sin \delta }{r_{e}}\right) , & 
\left\{ 
\begin{array}{c}
-\frac{\pi }{2}\leq \varphi \leq \phi _{-} \\ 
\phi _{+}\leq \varphi \leq \frac{\pi }{2}%
\end{array}%
\right. 
\end{array}%
\right. 
\end{equation}%
to cover the whole semicircular path from $\varphi =-\pi /2$\ to $\varphi
=\pi /2$. In these relations, $\phi _{+}$\ ($\phi _{-}$) denotes a special
point where $d\delta /d\varphi $\ vanishes for $\varphi >0$ ($\varphi <0$), as indicated in Fig. \ref{delatVSphiPlot}. 

\end{document}